\documentclass[a4paper]{jpconf}
\usepackage{graphicx}
\usepackage{iopams}
\usepackage{aas_macros}
\begin{document}
\def\teff{$T\rm_{eff }$}
\def\kms{$\mathrm {km s}^{-1}$}
\def\msun{\rm M_{\odot}}
\def\mbh{M_{\rm BH}}
\def\etal{{et al.\ }}
\def\simlt{\mathrel{\rlap{\lower 3pt\hbox{$\sim$}}\raise 2.0pt\hbox{$<$}}}
\def\simgt{\mathrel{\rlap{\lower 3pt\hbox{$\sim$}} \raise 2.0pt\hbox{$>$}}}
\def\lsim{\mathrel{\rlap{\lower 3pt\hbox{$\sim$}}\raise 2.0pt\hbox{$<$}}}
\def\gsim{\mathrel{\rlap{\lower 3pt\hbox{$\sim$}} \raise 2.0pt\hbox{$>$}}}
\def\di{\mbox{d}}
\def\mbulge{M_{\rm Bulge}}
\def\msunpc3{\msun~{\rm {pc^{-3}}}}
\newcommand{\be}{\begin{equation}}
\newcommand{\ee}{\end{equation}}

\title{Massive black hole binary evolution\\in gas-rich mergers}

\author{M. Colpi$^1$, S. Callegari$^2$, M. Dotti$^3$, and L. Mayer$^{2,4}$}

\address{$^1$ Department of Physics G. Occhialini, University
  of Milano Bicocca, Milano, Italy \\ $^{2}$ Insitute of Theoretical
  Physics, Zurich, Switzerland \\$^{3}$ Department of Astronomy, University of  Michigan, Ann Arbor, USA \\
  $^{4}$ Institute of Astronomy, Department of Physics, ETH, Zurich, Switzerland}

\ead{monica.colpi@mib.infn.it}

\begin{abstract}
We report on key studies on the dynamics of black holes (BHs) 
in {\it gas-rich} galaxy mergers that underscore the vital role played by 
gas dissipation in promoting BH inspiral down to the smallest scales ever probed
with use of high-resolution numerical simulations.
In  major mergers, the BHs 
sink rapidly under the action of gas-dynamical friction while 
orbiting inside the massive nuclear disc resulting from 
the merger.  The BHs then bind and form  
a Keplerian binary on a scale of $\lsim 5$ pc. In minor mergers, BH pairing proceeds down to the minimum scale explored of 10-100 pc  
only when the gas fraction 
in the less massive galaxy is comparatively large to avoid
its tidal and/or ram pressure disruption and the wandering of the light BH
in the periphery of the main halo. 
Binary BHs enter the gravitational wave
dominated inspiral only when their relative distance is typically of $\sim 10^{-3}$pc.
If the gas preserves the degree of dissipation
expected in a star-burst environment, 
binary decay continues down to 0.1pc, the smallest length-scale ever attained. 
Stalling versus hardening below $\lsim 0.1$pc  
is still matter of deep investigations, and there is no
unique answer depending on the yet unexplored dynamics of gas 
in the vicinity of the binary.
\end{abstract}

\section{Introduction}

Dormant black holes (BHs) with masses in excess of $\simgt 10^6\msun$
are found to be ubiquitous in bright spheroids, today \cite{kormendy95,richstone98}.
This local population comprises the dead remnants of a bright past, when 
the same BHs were powering the most luminous quasars. The
recent discovery of scale relations between the BH mass and the properties
of the stellar bulge \cite{ferrarese05}, likely set via AGN feedback, has prompted 
the study of BH growth and evolution in the general framework 
of galaxy evolution. According to the current $\Lambda$CDM paradigm for structure formation, galaxies
interact and merge as their dark matter halos assemble in a
hierarchical fashion \cite{springel06}, and BHs
incorporated through mergers into larger and more massive systems evolve concordantly: major central gas-inflows are triggered during the violence of a 
merger that feed the BH and power AGN activity \cite{hopkins08}.  
{\it In this context, close BH pairs form as
inescapable outcome of  galaxy evolution} \cite{kazantzidis05}.
In our local universe, NGC 6240 and Mrk 463 provide compelling evidence of 
ongoing gas-rich  mergers where two active nuclei are present, still
at large separations
of $\sim$ kpc.
Whether these BHs will spiral inward, form a  BH {\it binary}
and {\it coalesce} under the emission of
gravitational waves is a matter of our concern. The {\it Laser
Interferometer Space Antenna} ({\it LISA}) is expected to record
these extraordinary events 
out to redshift $z\sim 20$ providing  not only a firm
test of General Relativity, but also a view, albeit indirect, of galaxy
clustering together with an extremely accurate measure of the
BH mass and spin 
\cite{vecchio04,volonteri03,sesana05}.

Galaxy mergers
cover cosmological volumes (of hundred kpc aside) whereas BH
coalescences  probe volumes from a few parsecs (when they  bind in nuclear discs) down
to an astronomical unit and less.
Thus, following a merger, how can BHs reach the gravitational wave inspiral
regime?  
Our aim is at studying the BH dynamics in gas-rich environments, and in particular, 
the transit from the state P of {\it pairing}  when each BH moves individually inside the time-varying potential of the colliding galaxies, to state B  when the two BHs
dynamically couple their motion to form a {\it binary}.  After all
transient inflows have subsided and the 
new galaxy has formed, the BH binary, 
surrounded by a massive circum-nuclear disc,  enters phase H 
where it hardens 
to smaller separations under the action of gas-dynamical and gravitational torques,
ideally down to $\sim 10^{-3}$ pc (for typical BH
mass of $10^6\,\msun$) where the gravitational waves 
domain G starts.
There is a number of key questions to address:  
(i) How does transition from state P $\to$ B depend on the gas 
thermodynamics and level of dissipation? 
(ii) In the grand nuclear disc inside the remnant galaxy, how do orbits evolve?
(iii) Do the BHs reach the gravitational wave driven domain?
(iv) During the hardening through phase B $\to$ H, do the BHs   
collect substantial amounts of gas to form cold individual discs?

\section{Pairing of Massive Black Holes in gas-rich mergers}
There are two types of mergers: major mergers  between galaxies of comparable mass 
(1:1 mass ratio), and  minor mergers between galaxies with smaller mass ratios 
(1:10 typically). 

{\sl Major Mergers} has been studied with N-Body/SPH
simulations with unprecedented force
resolution (down to $\sim 1$ pc using splitting techniques and {\it GASOLINE}
as integrator; see \cite{mayer07})
%carried out  
%with the N-Body/SPH code {\sl Gasoline} (Wadsley,
%Stadel \& Quinn 2004)  
to describe the collision of two galaxies similar to
the Milky Way. Each galaxy comprises a central BH of $2.6\times 10^6\,\msun$, 
a stellar bulge, a disc of stars and gas (with mass fraction of 10\% relative
to the total disc mass),
and an extended spherical dark matter halo (of $10^{12}\,\msun$) with
NFW density profile (see \cite{mayer07} for details).

\begin{figure}
\begin{center}
\includegraphics[width=0.70\textwidth]{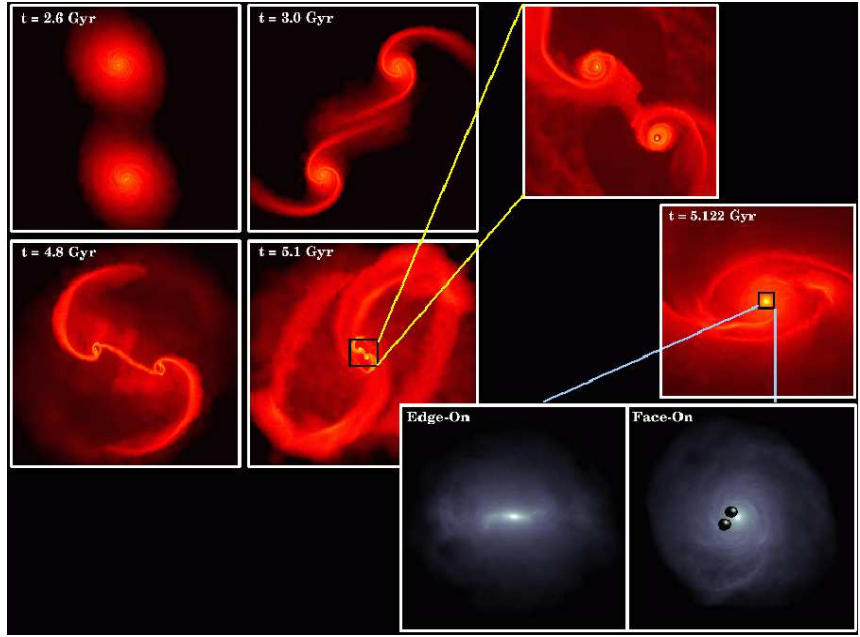}
\end{center}
\caption[]{
  The different stages of the merger between two identical disc
  galaxies seen face--on. The color-coded density maps of the gas
  component are shown using a logarithmic scale, with brighter colors
  for higher densities.  The four panels to the left show the
  large-scale evolution at different times (obtained with a force
  resolution of 100 pc). The boxes are 120 kpc on a
  side (top) and 60 kpc on a side (bottom) and the density ranges
  between $10^{-2}$ atoms cm$^{-3}$ and $10^{2}$ atoms cm$^{-3}$.
  During the interaction, tidal forces tear the galactic discs apart,
  generating spectacular tidal tails and plumes.  The upper panel to the
  right shows a zoom in view of the two discs before
  they merge into a single rotating nuclear gaseous disc embedded in a series of
  large-scale ring-like structures (middle panel).  The boxes are now 8
  kpc on a side and the density ranges between $10^{-2}$ atoms
  cm$^{-3}$ and $10^{5}$ atoms cm$^{ -3}$.  The two bottom panels,
  with a gray color scale, show the detail of the inner 160 pc of
  the middle panel (here the force resolution is 2 pc); 
  the nuclear disc is shown edge-on (left) and
  face-on (right), and the two BHs are also shown in the face-on
  image.}
\end{figure}

The galaxies first experience two close fly-bys:
in this early phase, the cuspy potentials of both
galaxies are deep enough to allow for the survival of their baryonic
cores that  sink under the action of dynamical friction
against the dark matter background, dragging together
the two BHs.  Strong spiral patterns
appear in both the stellar and gaseous discs, and as the
merger continues,  non--axisymmetric
torques redistribute angular momentum: as much as 60\% of the gas
originally present in each disc of the parent galaxies is funneled
inside the inner few hundred parsecs of the individual cores.  This is
illustrated in the upper right panel of Figure 1, where the enlarged
color coded density map of the gas is shown, after 5.1 Gyr from the
onset of the collision. {\it Each BH is surrounded by a
rotating stellar and gaseous disc of mass $\sim 4 \times 10^8$
M$_{\odot}$ and size of a few hundred parsecs.}  The two discs and BHs
are just 6 kpc far apart, and at the same time a star-burst of $\sim 30 \msun\,$yr$^{-1}$
has invested the central region of the ongoing merger.  

At this stage, the simulation is stopped and restarts with increased
resolution (of $\sim 2$ pc). In order to simulate the environment of a star burst where cool gas
coexists with the warm phase heated by stellar feedback,
the pressure is set equal to $P=(\gamma-1)\rho u$ with
$\gamma=7/5$ (according to fits by  \cite{spaans00}). 
The internal energy per particle $u$ evolves with time as a result of $P dV$ 
work and shock heating modeled via the standard Monaghan artificial viscosity term.

With time, the two baryonic discs get closer and closer and 
eventually merge in {\it a single
massive self-gravitating, rotationally supported 
nuclear disc}, now weighing $3\times 10^9\,\msun$. 
This is illustrated again in Figure 1 (mid and bottom right panels). 
The gaseous disc, dominant in mass, is surrounded by 
a background of dark matter and stars
distributed in a spheroid. 

\begin{figure}
\begin{center}
\includegraphics[width=0.48\textwidth]{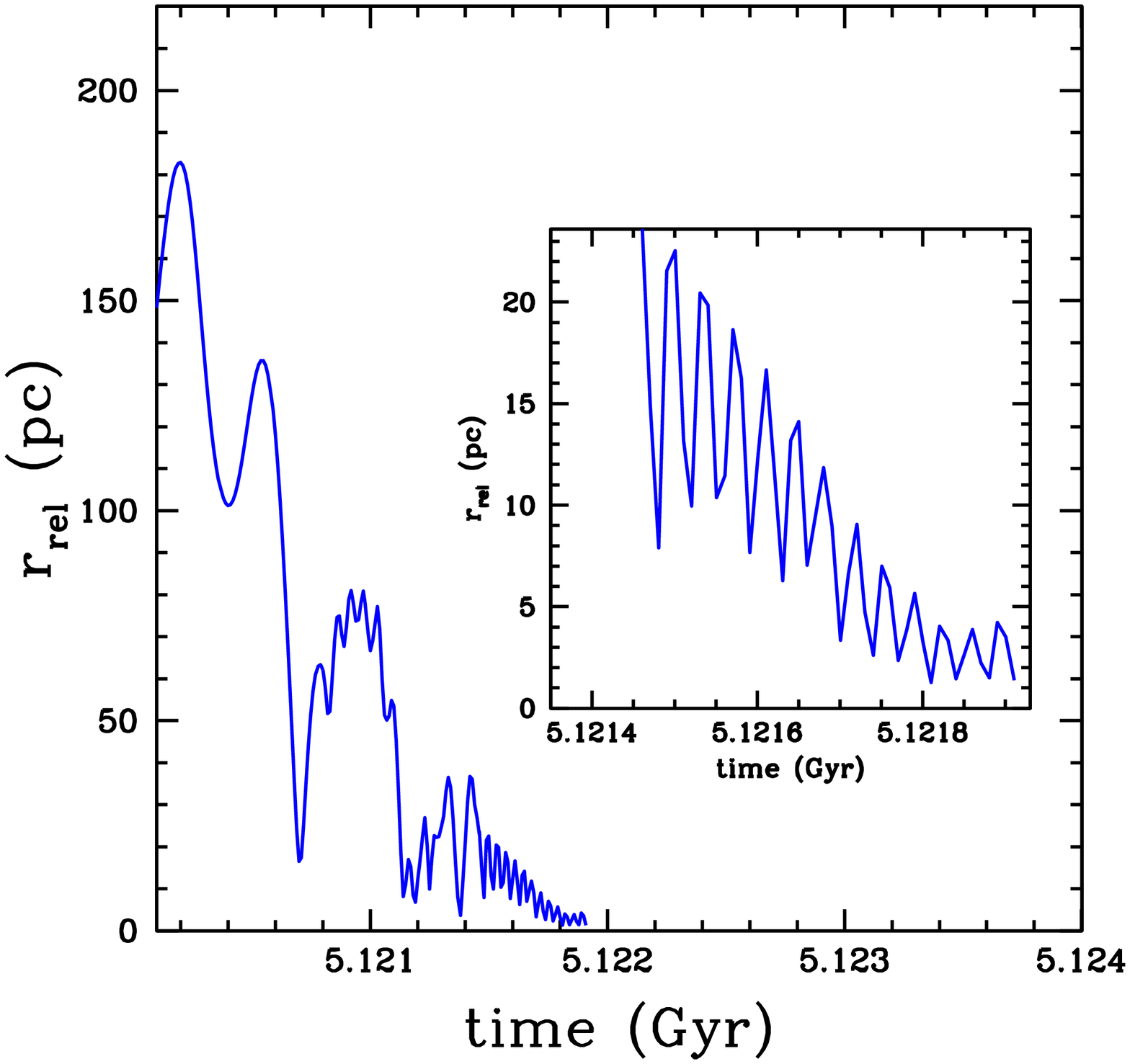}
\end{center}
\caption[]{Orbital separation of the two BHs as a function 
  of time during the last stage of the galaxy merger.  
  The orbit of the pair is eccentric until the end of the
  simulation. The two peaks at scales of tens of parsecs at around
  $t=5.1213$ Gyr mark the end of the phase during which the two holes
  are still embedded in two distinct gaseous cores. Until this point
  the orbit is the result of the relative motion of the cores combined
  with the relative motion of each BH relative to the surrounding
  core, explaining the presence of more than one orbital frequency.
  The inset shows the details of the last part of the orbital
  evolution, which takes place in the nuclear disc arising from the
  merger of the two cores. The binary stops shrinking when the
  separation approaches the force resolution limit (2 pc).}
\label{fig:birth}
\end{figure}

The BHs have been dragged together toward the
dynamical center of the merging galaxies,  and move inside the grand
disc spiraling inward under the action of gas-dynamical friction. 
In less than a million years after the merger, 
they eventually bind  gravitationally to each other, as the
mass of the gas enclosed within their separation is less than the mass
of the binary. It is the gas that controls the orbital decay, not the stars. 
The transition between state P to B is now completed as illustrated in
Figure 2.  
Dynamical friction against the stellar background would bring the two
BHs this close only on a longer timescale, $\sim 
10^8$ yr \cite{mayer07}.
This short 
sinking timescale 
comes from 
the combination of the fact that gas densities are much higher
than stellar densities in the center, and that in the
mildly supersonic regime the drag against a gaseous background is stronger than
that in a stellar background with the same density \cite{ostriker99}.
It is worth noticing that the transition P $\to $ B is sensitive to the
gas  thermodynamics: BH coupling is delayed 
if gas were to follow thermal evolution with a $\gamma=5/3$ \cite{mayer07}.

\begin{figure} 
\begin{center}
\includegraphics[width=0.7\textwidth]{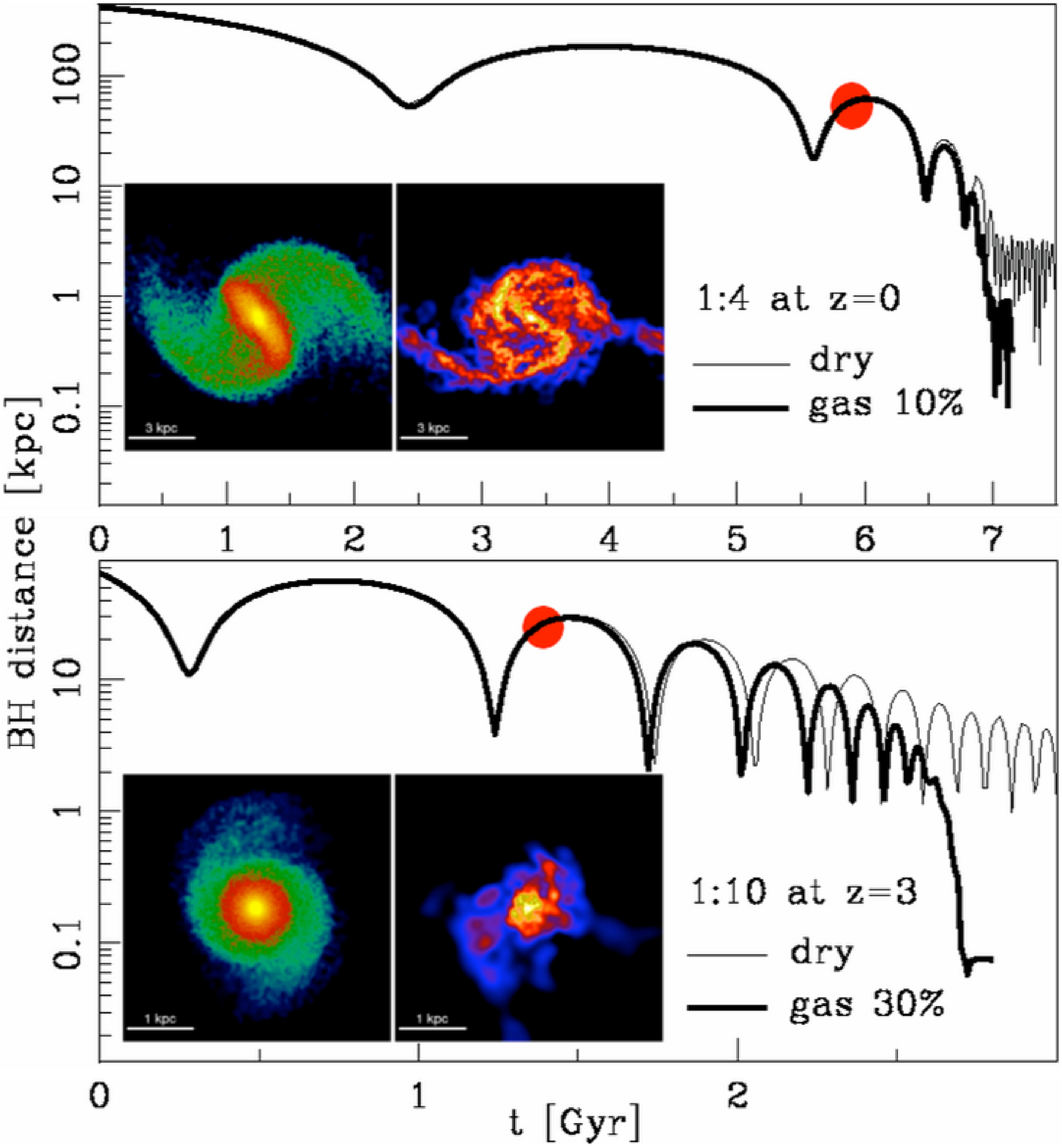}
\end{center}
\caption{
BH separation as a function of time in four of our simulations.  Upper row: 
BH distance in 1:4 mergers (for galaxy models at $z=0$); the thin and thick lines refer to simulations with 
no-gas and with gas ($f_{\rm gas}=0.1\%$), respectively.    
Lower row: BH distance for the 1:10 mergers (for galaxy models at $z=3$); the thin and thick lines refer to simulations with 
no-gas and with gas ($f_{\rm gas}=0.3\%$), respectively.    
The insets  show the color-coded density maps 
of stars (left) and gas (right), 4 kpc on a side.
The large dot on the BH curve indicates the time at which
the two snapshots are recorded.  Colors code the range
  $10^{-2}-1$~M$_\odot$~pc$^{-3}$ for stars, and
  $10^{-3}-0^{-1}$~M$_\odot$~pc$^{-3}$ for the gas.
  These snapshots are representative of the average behavior of the
  satellites during the first two orbits.
  Note the formation of a strong bar for the 1:4 minor merger, which is absent for
  the 1:10 case, and the truncation of the gaseous disc in the 1:10 
  satellite caused by ram pressure stripping. 
  \label{fig:minor}}
\end{figure}

Does BH pairing proceed similarly, in {\sl minor mergers} predicted to
be common events in the high redshift universe \cite{volonteri03}
and of primary importance for {\it LISA} \cite{sesana05}?        
To answer this question we extended our numerical investigation 
to 1:4 mergers at 
$z=0$, and 1:10 mergers at $z=3$, assuming a roughly constant 
$M_{\rm BH}-M_{\rm bulge}$ relation in between these cosmic
epochs, and initial galaxy models   
replica of a Milky Way suitably rescaled in mass and size
(see for details \cite{callegari08}. The masses of the two BHs in the $z=3$ runs 
are thus $6\times10^5$ and $6\times10^4M_\odot$, and their expected
inspiral and coalescence signal falls nicely in the {\it LISA} sensitivity
window \cite{sesana05}.
 
It is found that  minor mergers differ profoundly from major mergers
as early noticed by  \cite{governato94}. The
encounter is closer to an {\it accretion} process whereby the 
less massive galaxy is dramatically damaged during its sinking into the primary.
In our recent study  \cite{callegari08}, paring is found to
be very sensitive to the details of the
physical processes involved. In all cases with no-gas (i.e., in "dry" runs)
the formation of a close BH pair is aborted: tidal shocks progressively
lower the density in the satellite until it dissolves, leaving a
wandering black hole in the remnant.  This is illustrated in Figure 3 (thin lines)
where the BH relative distance remains as large as 1-10 kpc. 
Only with the inclusion of
a cold gaseous disc component, and star formation the outcome of
the merger changes significantly.

Figure 3 depicts the   
the stellar and gaseous components of the satellite to show
their  profound structural damage (the primary is not shown).  
\begin{figure}
\begin{center}
\includegraphics[width=0.48\textwidth]{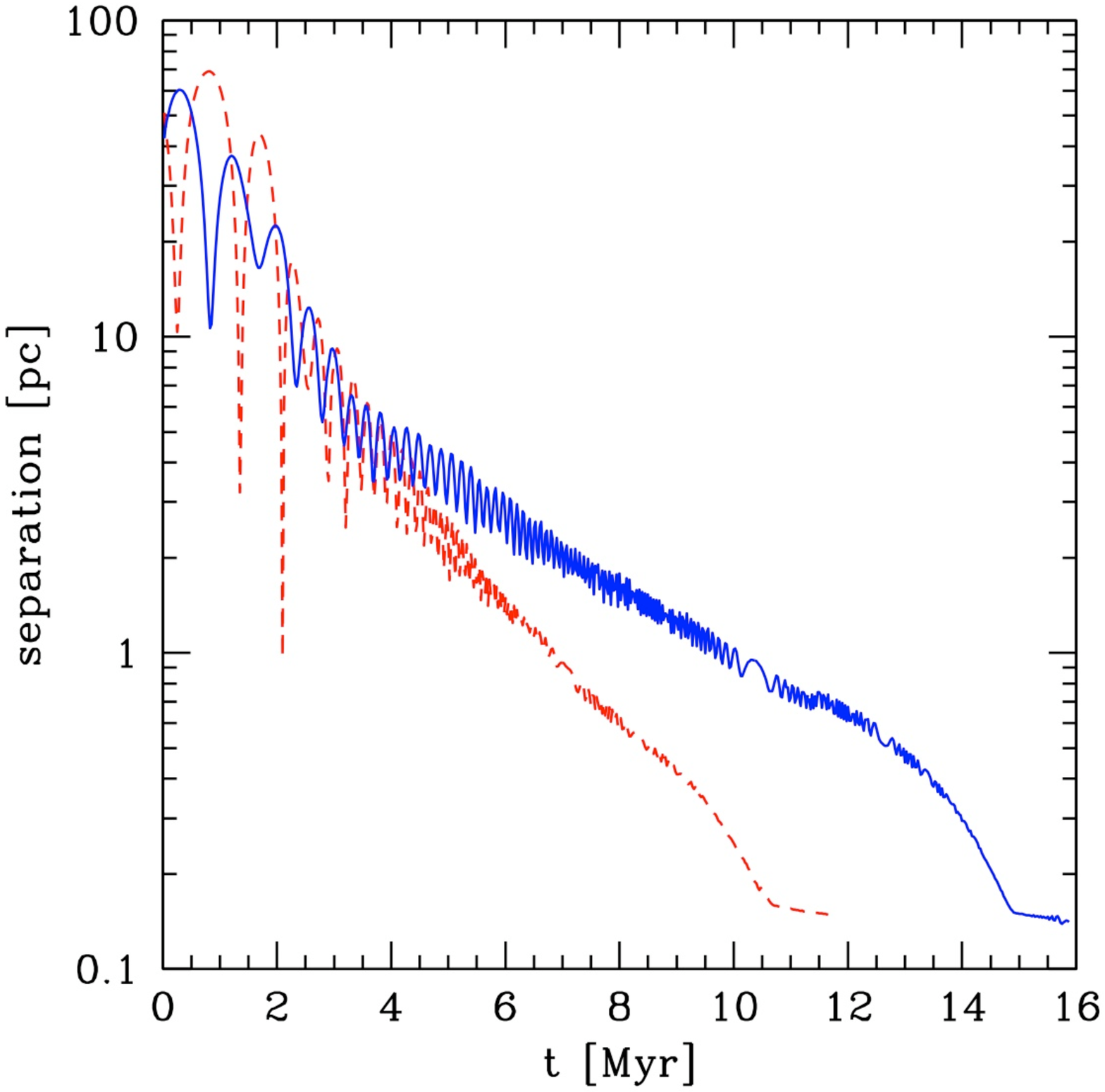}
\end{center}
\caption[]{ BH separation as a function of time. 
One BH is set at the center of a rotationally supported disc while 
the secondary BH moves initially on an eccentric ($e=0.7$) 
either co- or counter-rotating orbit.
BH masses are of $10^6\,\msun$,. The  gaseous disc 
of mass $M_{\rm disc}=10^8\,\msun$ is described by a Mestel profile.
Solid (blue) line refers to the co-rotating case, while (red) dashed line
  refers to the counter-rotating case (Dotti et al. in preparation).  }
\label{fig:birth}
\end{figure}
%
%Thanks to the high spatial resolution of our simulations that allows.
For mass ratios 1:4 at $z=0$,
bar instabilities excited at pericentric passages funnel gas 
(present in a fraction $f_{\rm gas}$ of the total disc mass) to the
center of the 
satellite, steepening its potential well and allowing its survival
against tidal disruption down to the center of the merger remnant. 
As shown in Figure 3 (thick lines), the BHs pair down to $\sim100$~pc scales
(the force resolution limit),
creating conditions favorable to the formation of a BH binary. 
The smaller satellites (with 1:10 mass ratio
at $z=3$) are more strongly 
affected by both internal star formation and the gas-dynamical
interaction between their interstellar medium and that of the primary galaxy. 
Torques in the early stages of the merger are not acting to concentrate gas to
the center, due to the absence of a stellar bar and the
stabilizing effect of 
turbulence. As a result, ram pressure strips all of the ISM of the
satellite. Only the gas-rich satellites (those with $f_{\rm gas}=0.3$)
undergo a central burst of star formation during the first orbits which  
increases their central stellar density allowing for their
survival. In these models, pairing of the two BHs via dynamical
friction occurs down to $\sim 100$ pc, a few Gyrs after the disruption of the satellite
(see Figure 3).
 
\begin{figure}
\begin{center}
\includegraphics[width=0.40\textwidth]{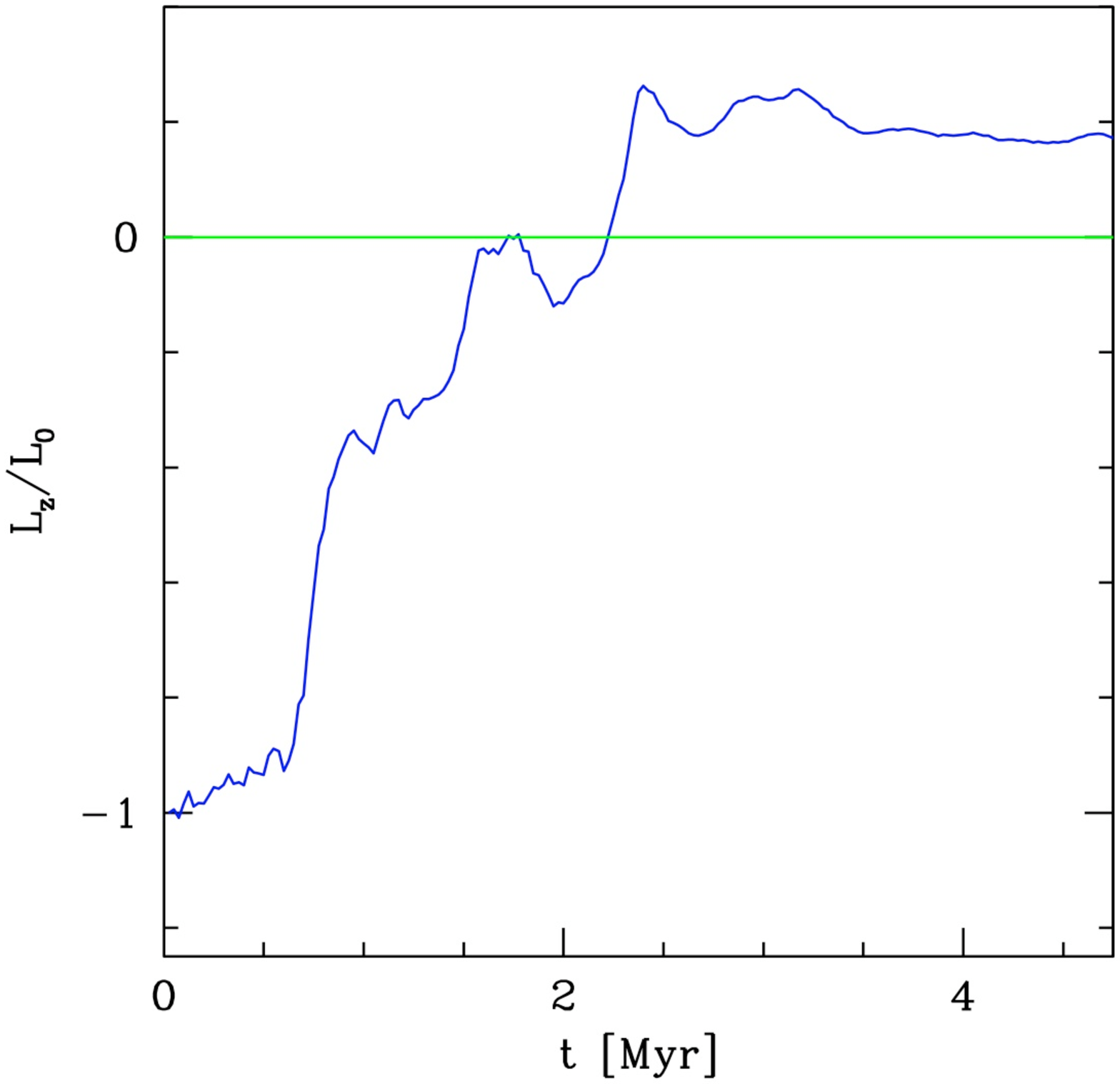}
\end{center}
\caption[]{$z$-component of the orbital angular momentum of the secondary BH
normalized to its initial value $L_0$ for the counter-rotating case,  
described also in Figure 4, to show the occurrence of the angular 
momentum flip induced by dynamical friction \cite{dotti09}.}
\label{fig:birth}
\end{figure}

\section{Black hole binaries in massive nuclear discs}

As shown in Section 2, massive circum-nuclear discs form in the aftermath
of a major gas-rich merger. It is in these discs  that the BHs complete their
transition from P$\to$ B, and continue to spiral inward under the
action of gas-dynamical torques from B $\to$ H
\cite{escala05,dotti06,dotti07}. 
A still open issue is whether the BHs will reach the domain of 
gravitational waves  inspiral  within a Hubble time: for a
$10^{6}\,\msun$ BH binary on a circular orbit, the
transition H $\to$ G occurs when the separation is 
around $10^{-3}$ pc. 
Can material and gravitational torques be effective in driving the BHs down to this tiny scale?

Here we describe our attempts to 
explore the transition from P $\to $ B $\to$ H, 
for BHs orbiting inside a circum-nuclear disc using {\it GADGET} as N-Body/SPH
code with a force resolution of only  $\approx 0.1$ pc, and no
splitting during the entire course of the evolution \cite{dotti09}. 
In our selected model, a BH (called primary) is set at the
center of a massive differentially rotating gaseous disc in equilibrium with a stellar bulge
(see \cite{dotti06,dotti07} for the details).  
A second BH 
(called secondary) with similar/equal mass is delivered
at a large distance ($\sim 50$ pc) from the center, along a coplanar 
eccentric ($e=0.7$) orbit that can either be co- or counter-rotating
relative to the background disc.
The $10^6$ SPH particles, making the disc, evolve 
as in \cite{mayer07}: accordingly, the gas thermodynamics is described by 
the index $\gamma= 7/5$
that accounts for the presence of net cooling in a star-forming region. Shocks
here are less important as the equilibrium disc is only mildly perturbed by the 
BHs.
Figure 4 shows the BH separation 
as a function of time for the co-rotating and counter-rotating cases. 
No stalling is observed in both cases, as the relative BH distance decays 
rapidly (in $\lsim 20$ Myr) down
to the force resolution length-scale. In the final stages, 
the more rapid orbital decay is due to
the torque exerted by the ellipsoidal deformation 
that forms in the gas when the heads of the density wakes overlap and whose axis is misaligned relative to the BH binary axis \cite{escala05}. 

In the counter-rotating case, the angular
momentum of the secondary BH (initially negative) grows very efficiently during
the first Myr, when the BH is passing through the central, high density
region of the disc.  Angular momentum continues to grow monotonically
for the next $3-4$ Myrs, then becomes positive, i.e. the BH starts
to move on a co-rotating orbit with respect to the disc.
This is illustrated in Figure 5.
It is  dynamical friction that causes this orbital ``angular momentum flip''.
In both cases the eccentricity
of the orbit decreases to very small values (also in the counter-rotating case, after 
the orbit becomes co-rotating) due to the different response of the fluid to
the gravitational pull of the BH at the different orbital phases. 
When at pericenter the BH
moves faster than the gas and it is decelerated by the density wake excited behind its trail; when at apocenter the BH is moving more slowly than the gas and the wake is trailing in front causing a tangential acceleration. The composite effect is
a decrease of $e$.

In a suite of runs, the BHs have been 
modeled as ``sink particles'', i.e. they are allowed to accrete gas particles
during their dynamical evolution \cite{dotti09}.  We introduced an "on flight"
algorithm for accretion and determined the amount of
gas that binds to the BHs. 
It is only when the BH binary circularizes that gas is accreted to such an extent that
both BHs are surrounded by their own {\it accretion} disc, and these discs are expected
to play a role in guiding the subsequent hardening phase down to the gravitational domain.

\section{Open issues}
Numerical simulations, carried on with unprecedented accuracy, have revealed that 
the transit of dual BHs from P $\to$ B $\to$ H, and finally from H$\to$ G
is a sensitive function of the merger type and of amount of cold gas present in the 
interacting galaxies.
While the transition from P (pairing) $\to$ B (binary formation) appears to be 
likely in gas-rich major mergers as well as in gas-rich minor mergers
at high redshift, hardening  down to the
gravitational wave domain remains still uncertain
on scales below $\sim 0.1$ pc  and not fully explored. 
It has been suggested that a circum-binary viscous disc 
inevitably forms around
the BH binary on sub-parsec scales that absorbs the
the angular momentum of the binary \cite{cuadra08}. 
This circum-binary disc would represent the last cold 
environment for BH hardening from H $\to $ G. 
Braking of the BH rapid motion 
requires energy loss and angular momentum transport 
trough a mechanism that is reminiscent of planet migration in
proto-stellar discs {\cite{cuadra08,gould00}: while tidal torques from the BH binary 
carry away orbital angular momentum, viscous torques inside the 
disc sustain the radial motion of the gas 
toward the BHs, 
maintaining the binary in near contact with the disc. 
Equilibrium between these two torques would cause the slow drift of the
BHs toward smaller and smaller separations, until gravitational
waves guide the final inspiral.  No calculation has reproduced yet the 
formation of a circum-binary disc from the earlier phase, 
in a self-consistent manner, nor it is clear how fast 
will be the inspiral, and how large the growth of 
the eccentricity \cite{armitage02}.  The BH binary likely enters 
phase G with a residual  
eccentricity still imprinted in the gravitational wave signal, despite
the circularizing action of the gravitational wave back reaction.
{\it LISA} is expected to be lunched by 2020. 
By that time, our theoretical understanding of
binary hardening in a gas-rich environment
will hopefully improve thanks to the progress expected in numerical simulations, 
and in our ability to model fragmentation, star formation and feedback, 
inside galactic nuclei.

\medskip
We thank all collaborators that made this research possible: F. Governato,
S. Kazantzidis, F. Haardt, 
P. Madau, B. Moore, L. Paredi, J. Wadsley, M. Ruszkowski, J. Stadel, T. Quinn,  and M. Volonteri. 
\section*{References}
\bibliography{biblio}{}

\providecommand{\newblock}{}
\begin{thebibliography}{10}
\expandafter\ifx\csname url\endcsname\relax
  \def\url#1{{\tt #1}}\fi
\expandafter\ifx\csname urlprefix\endcsname\relax\def\urlprefix{URL }\fi
\providecommand{\eprint}[2][]{\url{#2}}

\bibitem{kormendy95}
{Kormendy} J and {Richstone} D 1995 {\em \araa\/} {\bf 33} 581--+

\bibitem{richstone98}
{Richstone} D, {Ajhar} E~A, {Bender} R, {Bower} G, {Dressler} A, {Faber} S~M,
  {Filippenko} A~V, {Gebhardt} K, {Green} R, {Ho} L~C, {Kormendy} J, {Lauer}
  T~R, {Magorrian} J and {Tremaine} S 1998 {\em \nat\/} {\bf 395} A14+

\bibitem{ferrarese05}
{Ferrarese} L and {Ford} H 2005 {\em Space Science Reviews\/} {\bf 116}
  523--624 (\textit{Preprint} \eprint{arXiv:astro-ph/0411247})

\bibitem{springel06}
{Springel} V, {Frenk} C~S and {White} S~D~M 2006 {\em \nat\/} {\bf 440}
  1137--1144

\bibitem{hopkins08}
{Hopkins} P~F, {Hernquist} L, {Cox} T~J and {Kere{\v s}} D 2008 {\em \apjs\/}
  {\bf 175} 356--389 (\textit{Preprint} \eprint{0706.1243})

\bibitem{kazantzidis05}
{Kazantzidis} S, {Mayer} L, {Colpi} M, {Madau} P, {Debattista} V~P, {Wadsley}
  J, {Stadel} J, {Quinn} T and {Moore} B 2005 {\em \apjl\/} {\bf 623} L67--L70

\bibitem{vecchio04}
{Vecchio} A 2004 {\em \prd\/} {\bf 70} 042001--+

\bibitem{volonteri03}
{Volonteri} M, {Haardt} F and {Madau} P 2003 {\em \apj\/} {\bf 582} 559--573

\bibitem{sesana05}
{Sesana} A, {Haardt} F, {Madau} P and {Volonteri} M 2005 {\em \apj\/} {\bf 623}
  23--30

\bibitem{mayer07}
{Mayer} L, {Kazantzidis} S, {Madau} P, {Colpi} M, {Quinn} T and {Wadsley} J
  2007 {\em Science\/} {\bf 316} 1874--

\bibitem{spaans00}
{Spaans} M and {Silk} J 2000 {\em \apj\/} {\bf 538} 115--120 (\textit{Preprint}
  \eprint{arXiv:astro-ph/0002483})

\bibitem{ostriker99}
{Ostriker} E~C 1999 {\em \apj\/} {\bf 513} 252--258 (\textit{Preprint}
  \eprint{arXiv:astro-ph/9810324})

\bibitem{callegari08}
{Callegari} S, {Mayer} L, {Kazantzidis} S, {Colpi} M, {Governato} F, {Quinn} T
  and {Wadsley} J 2008 {\em ArXiv e-prints\/} (\textit{Preprint}
  \eprint{0811.0615})

\bibitem{governato94}
{Governato} F, {Colpi} M and {Maraschi} L 1994 {\em \mnras\/} {\bf 271} 317--+

\bibitem{dotti09}
{Dotti} M, {Ruszkowski} M, {Paredi} L, {Colpi} M, {Volonteri} M and {Haardt} F
  2009 {\em ArXiv e-prints\/} (\textit{Preprint} \eprint{0902.1525})

\bibitem{escala05}
{Escala} A, {Larson} R~B, {Coppi} P~S and {Mardones} D 2005 {\em \apj\/} {\bf
  630} 152--166

\bibitem{dotti06}
{Dotti} M, {Colpi} M and {Haardt} F 2006 {\em \mnras\/} {\bf 367} 103--112

\bibitem{dotti07}
{Dotti} M, {Colpi} M, {Haardt} F and {Mayer} L 2007 {\em \mnras\/} {\bf 379}
  956--962

\bibitem{cuadra08}
{Cuadra} J, {Armitage} P~J, {Alexander} R~D and {Begelman} M~C 2008 {\em ArXiv
  e-prints\/} (\textit{Preprint} \eprint{0809.0311})

\bibitem{gould00}
{Gould} A and {Rix} H~W 2000 {\em \apjl\/} {\bf 532} L29--L32

\bibitem{armitage02}
{Armitage} P~J and {Natarajan} P 2002 {\em \apjl\/} {\bf 567} L9--L12
  (\textit{Preprint} \eprint{arXiv:astro-ph/0201318})

\end{thebibliography}

\end{document}